\documentclass[conference]{IEEEtran}
\IEEEoverridecommandlockouts

\usepackage{cite}
\usepackage{amsmath,amssymb,amsfonts}
\usepackage{algorithmic}
\usepackage{graphicx}
\usepackage{textcomp}
\usepackage{xcolor}
\def\BibTeX{{\rm B\kern-.05em{\sc i\kern-.025em b}\kern-.08em
    T\kern-.1667em\lower.7ex\hbox{E}\kern-.125emX}}

\begin{document}

\title{$M^3$ST-DTI: A Multi-Task Learning Model for Drug-Target Interactions Based on Multi-Modal Features and Multi-Stage Alignment}

\author{
\IEEEauthorblockN{Xiangyu Li}
\IEEEauthorblockA{College of Intelligence\\ and Computing\\Tianjin University\\Tianjin, China\\ Email: xiangyuli@tju.edu.cn}
\and
\IEEEauthorblockN{Ran Su}
\IEEEauthorblockA{College of Intelligence\\ and Computing\\Tianjin University\\Tianjin, China\\Email: ran.su@tju.edu.cn}
\and
\IEEEauthorblockN{Liangliang Liu*}
\IEEEauthorblockA{College of Information \\and Management Science\\ Henan Agricultural University\\Zhengzhou China\\Email: liangliu@henau.edu.cn}
}
\maketitle
\begin{abstract}
Accurate prediction of drug-target interactions (DTI) is pivotal in drug discovery. However, existing approaches often fail to capture deep intra-modal feature interactions or achieve effective cross-modal alignment, limiting predictive performance and generalization. To address these challenges, we propose $M^3$ST-DTI, a multi-task learning model that enables multi-stage integration and alignment of multi-modal features for DTI prediction. $M^3$ST-DTI incorporates three types of features-textual, structural, and functional and enhances intra-modal representations using self-attention mechanisms and a hybrid pooling graph attention module. For early-stage feature alignment and fusion, the model integrates MCA with Gram loss as a structural constraint. In the later stage, a BCA module captures fine-grained interactions between drugs and targets within each modality, while a deep orthogonal fusion module mitigates feature redundancy.Extensive evaluations on benchmark datasets demonstrate that $M^3$ST-DTI consistently outperforms state-of-the-art methods across diverse metrics.
\end{abstract}


\section{Introduction}
In the study of modern medicine, accurately predict drug - target interaction (DTI) has become a key foundation of the drug discovery and development process of \cite{chen2016drug,wen2017deep, chen2018machine}. However, as a result of the experiment method of the nature of the labor-intensive, time-consuming and expensive, large-scale still challenging for identification of the new type of DTI \cite{dimasi2003price,paul2010improve, Xu2025Redefining}. Therefore, the development of efficient and accurate computational methods for DTI prediction has become a task of strategic importance \cite{zhang2022deepmgt}.

In recent years, the method based on machine learning and deep learning in DTI prediction has attracted more and more attention \cite{yu2025ai,wang2023fusion, liu2025transformers}. These methods often use a type of input data, using machine learning, GNNs or based on sequence model (for example, CNN, RNN, transformer) to extract features and effectively modeling nonlinear relationship \cite{liu2020survey, gu2025artificial}. For example, Lee et al. \cite{lee2019deepconv} proposed DeepConv-DTI, which uses 1D-CNN to extract protein features and ECFP to encode drug structures, followed by prediction using a fully connected network. Huang et al. \cite{huang2021moltrans} introduced MolTrans to construct interpretable drug-target interaction maps to guide prediction. Zhao et al \cite{zhao2022hyperattentiondti} HyperAttentionDTI was proposed, combined with super attention mechanism to enhance characteristic expression ability. Tsubaki et al. \cite{tsubaki2019compound} used gnn for drug feature extraction and cnn for protein feature extraction. Cheng et al. \cite{cheng2022iifdti} proposed IIFDTI, which combines cnn and GATs in an encoder-decoder framework for drug-protein representation learning. Despite their success, single-modal methods face significant challenges, with most existing methods relying on source domain priors (such as Word2Vec \cite{chen2020transformercpi} or Graph neural networks \cite{ye2022molecular}) to build embedding representations. These methods often fail to generalize to new structures and semantics. This leads to noticeable performance degradation in cross-domain and cold-start scenarios. The advent of bioomics has stimulated DTI research, leading to the development of methods that exploit multimodal data. Hu et al. \cite{hu2025multi} leverage 1D, 2D, and 3D embeddings of drugs and proteins under the UnitedDTA framework to obtain a unified and discriminative representation of cross-modal alignments. Hua et al \cite{hua2025mmdg} proposed MMDG-DTI, which fuses generalized textual and structural features through domain adversarial training (DAT) and contrastive learning to enhance generalization ability. However, Du et al. \cite{du2022compound} points out that current DTI models still exploit the richness of multimodal data. The fusion strategy is still a key bottleneck: early fusion often ignores cross-modal interactions while building a comprehensive representation. Late fusion captures complementary signals, but may introduce redundancy and noise, affecting robustness and discrimination.

To address the challenges faced by DTI prediction, we propose $M^3$ST-DTI, a multi-modal, multi-stage feature fusion framework based on multi-task deep learning. The model extracts and integrates information from textual, structural and functional modalities. The main contributions of this study are as follows:
\begin{itemize}
\item[$\bullet$] We propose $M^3$ST-DTI, a multi-modal, multi-stage fusion, multi-task learning framework to construct textual, structural, and functional features from standard drug-target interaction datasets, significantly improving prediction accuracy and generalization ability.
\item[$\bullet$] First, early fusion is combined with Gram loss and multi-source cross-attention for semantic modality alignment, and deep orthogonal fusion is used to alleviate later redundancy.
\item[$\bullet$] We use self-attention and hybrid pooling graph attention to handle fine-grained feature patterns, and bidirectional cross-attention and hybrid graph convolution to handle robust drug-target interactions.
\item[$\bullet$] Extensive experiments verify the robustness of $M^3$ST-DTI, providing interpretability for precision drug design.
\end{itemize}
\section{Materials and Methods}
\subsection{Benchmark Datasets}
We used the public BindingDB dataset \cite{liu2007bindingdb} to train and evaluate our framework. The BindingDB dataset comprises 10,665 drugs and 1,413 target proteins, with 32,601 interactions measured by dissociation constant (Kd). Among these, 9,166 interactions are labeled as positive and 23,435 as negative. In total, we obtained 32,601 samples from BindingDB. To ensure a fair and consistent evaluation, we adopted the same data splits (training, validation, and testing) as those used in previous studies \cite{huang2021moltrans,kang2022fine}.
\subsection{Multi-modal Feature Construction}
\begin{figure}[htbp]
\centerline{\includegraphics[width=\columnwidth]{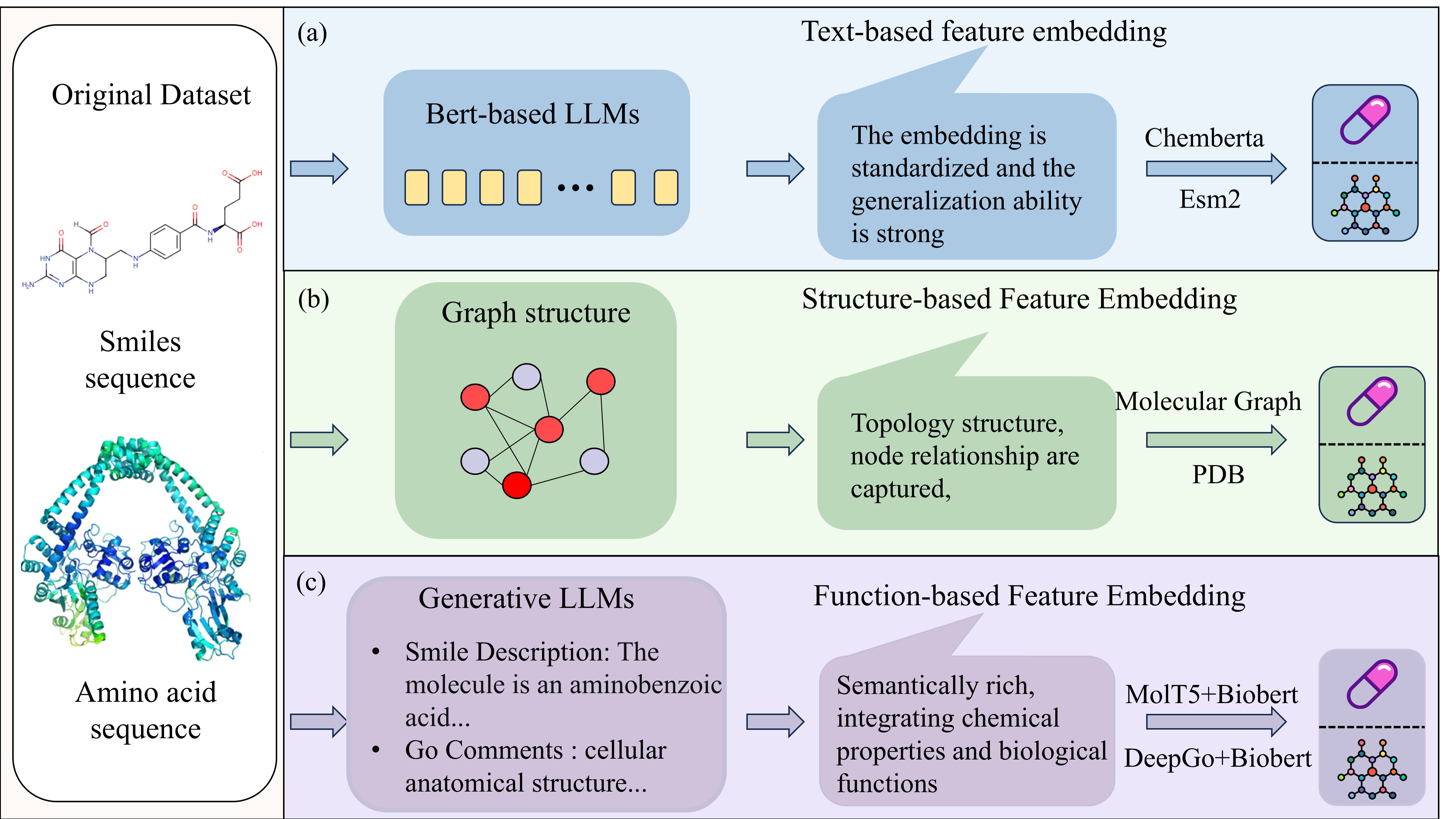}}
\caption{The process of constructing multi-modal features from a standard DTI dataset.(a) the process of constructing textual features (b) the process of constructing structural features (c) the process of constructing functional featuresn.}
\label{fig1}
\end{figure}

The proposed $M^3$ST-DTI framework constructs expressive and biologically meaningful representations for both drugs and protein targets by leveraging features from three complementary modalities: text, structure, and function, as shown in Fig.~\ref{fig1}. Feature extraction for each modality is performed using pre-trained models tailored to preserve domain knowledge and enhance modality-specific representations. 
\begin{figure*}[htbp]
\centering
\includegraphics[width=\textwidth,keepaspectratio]{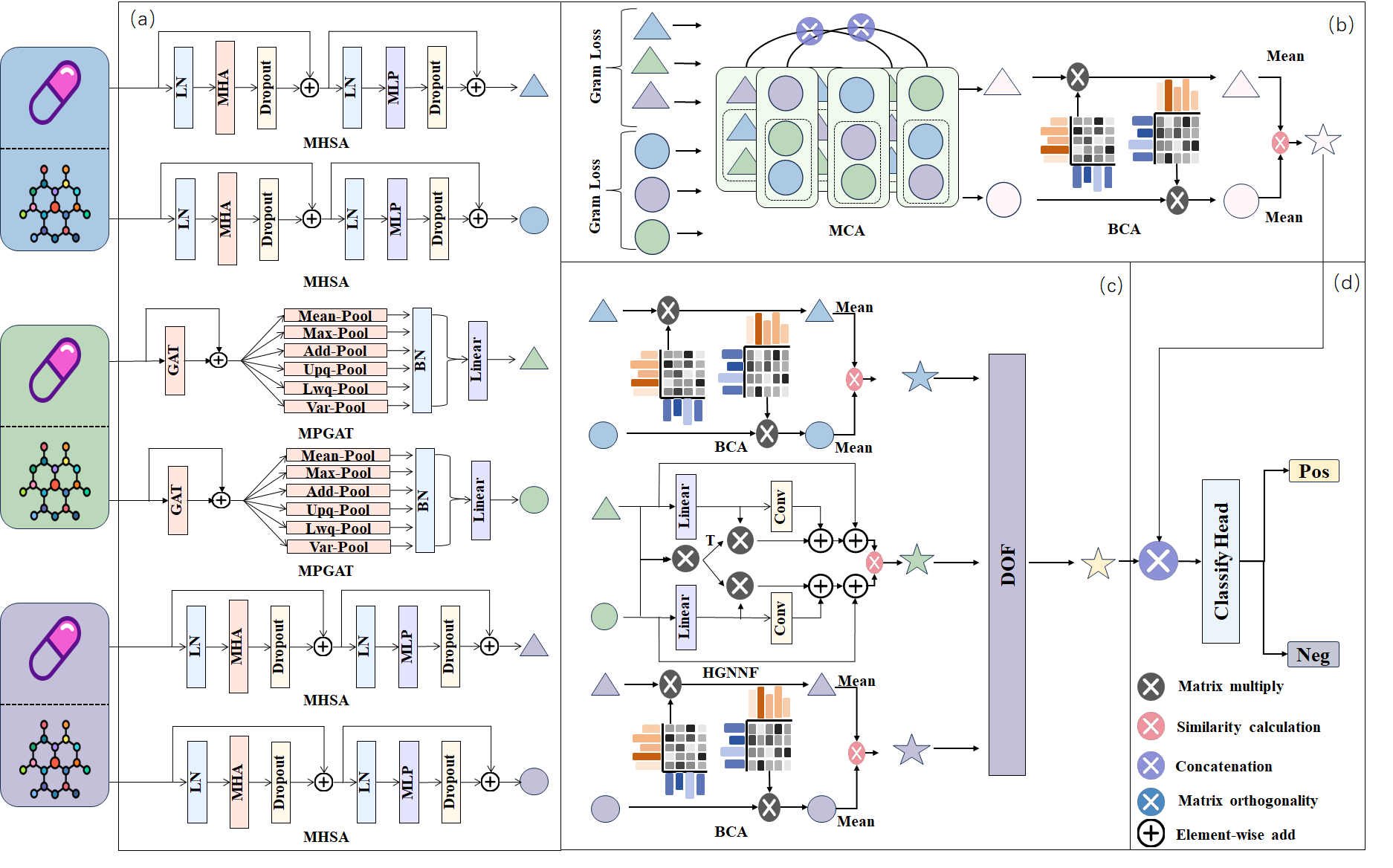}
\caption{Architecture flow of $M^3$ST-DTI method: (a) multi-modal feature extraction stage (b) early fusion stage (c) late fusion stage (d) interaction prediction stage.}
\label{fig2}
\end{figure*}

Based on these multi-modal extracted features, We propose a multi-task learning deep learning framework based on multi-modal data and multi-stage feature fusion ($M^3$ST-DTI), The process of $M^3$ST-DTI framework as shown in Fig.~\ref{fig2}

\subsection{Multi-modal feature extraction stage}
This section will introduce the screening process of text, structure and functional features. As shown in Fig. \ref{fig2}(a), each modal feature is processed independently to retain its domain-specific semantics.

\subsubsection{Text Feature Extraction}
As the foundational modality in DTI modeling, textual features describe drug SMILES strings and protein amino acid sequences with high-resolution semantic detail. To capture rich contextual dependencies, we apply a multi-head self-attention mechanism that enables each position in the sequence to attend to all others.
\subsubsection{Structure Feature Extraction}
Structural features are crucial in drug-target interaction prediction, capturing the graph topology of drug molecules and protein spatial conformations. We use graph-based structural modality and employ Graph Attention Networks (GATs) to dynamically weight nodes based on neighborhood relationships and edge attributes. To address limitations of global mean or max pooling, which may miss valuable information, we propose a Mixed Pooling Graph Attention (MPGAT) Module for comprehensive structural feature extraction from multiple perspectives.

\subsubsection{Function Feature Extraction}
Functional features in drug-target interaction prediction provide high-level semantic information regarding the biological processes, molecular functions, and cellular components associated with the molecules in biological systems. These features compensate for the limitations of structural and sequence features in capturing biological semantics. The feature extraction approach for the functional modality employed here is identical to the textual feature extraction method.

\subsection{Multi-level Feature Fusion}
\subsubsection{Early Fusion Stage}
As shown in Fig.~\ref{fig2}(b), the early fusion strategy first integrates multiple modal features-textual, structural, and functional-of drugs and proteins jointly, producing unified feature representations that span all three modalities. Subsequently, interactions between drug and protein features are modeled. This early fusion approach enhances feature completeness and consistency by leveraging synergistic interactions both within and across modalities. To achieve effective alignment among the three modalities, we introduce a loss function based on the Gramian Representation Alignment Measure Loss (GRAM Loss) \cite{cicchetti2024gramian}, which constrains the semantic closeness of different modal features in the embedding space. Set text, structure, and function of the modal characteristics of $x_{txt}^{'},x_{graph}^{'},x_{fun}^{'} \in {{\mathbb{R}}^{Seq \times Dim}}$, Gram Loss can be represented as:
\begin{equation}
\begin{aligned}
& x_{m}' = \frac{x_{m}'}{\sqrt{\sum_{j=1}^{\text{dim}} (x_{m,j}')^2}}, \quad m \in \{ \text{txt}, \text{graph}, \text{fun} \} \\
& G = \begin{pmatrix}
x_{txt}' \cdot x_{txt}' & x_{txt}' \cdot x_{graph}' & x_{txt}' \cdot x_{fun}' \\
x_{graph}' \cdot x_{txt}' & x_{graph}' \cdot x_{graph}' & x_{graph}' \cdot x_{fun}' \\
x_{fun}' \cdot x_{txt}' & x_{fun}' \cdot x_{graph}' & x_{fun}' \cdot x_{fun}'
\end{pmatrix}  \\
& V = \sqrt{\det(G) + \varepsilon} \\
& Gram Loss = -\frac{1}{B} \sum_{i=1}^{B} \log \left( \frac{\exp(-V_i / \tau)}{\sum_{j=1}^{k} \exp(-V_j / \tau)} \right)
\end{aligned}
\end{equation}
Where, $G\in {{\mathbb{R}}^{Seq\times 3\times 3}}$ represents the feature similarity matrix, $\det(\cdot)$ represents the volume calculation function, and $v \in {{\mathbb{R}}^{Seq \times 3\times 3}}$ represents the feature similarity matrix. $\varepsilon \text{=1}\times {{e}^{-8}}$ is a small constant to prevent numerical instability due to a determinant of zero, $V$ represents the volume of the parallel volume formed by the features, and $\tau$ is a temperature parameter.

In early fusion stage, we utilize a Multi-source Cross-attention (MCA) module to dynamically capture fine-grained interactions among the textual, structural, and functional modality features. MCA allows one modality (text) to dynamically focus on the relevant information of other modalities (structure, function), capturing the fine-grained interaction between modalities.

In addition, to ensure balanced representation of drug and protein features, we introduce a Bidirectional Cross-Attention (BCA) module that enables dynamic, two-way interaction between modalities. Unlike prior co-attention methods that risk overemphasizing drug features and weakening protein representations, our BCA module captures complex semantic associations through bidirectional information flow. This promotes semantic consistency and leads to comprehensive, unified embeddings for drug target interaction (DTI) prediction.
\subsubsection{Late Fusion Stage}
As shown in Fig.~\ref{fig2}(c), in the $M^3$ST-DTI framework, the core idea of the late fusion strategy is to perform deep intra-modal interactions. For each of the three modalities-textual, structural, and functional-we apply a BCA mechanism and a Hybrid Graph Convolutional Fusion (HGCNF) module to drug and protein features, respectively. These modules are designed to uncover hidden associations from multiple perspectives, yielding three sets of high-quality representations for drug-target interactions. Compared to the BCA module, the HGCNF module, tailored for the structural modality, integrates a 1D convolution operation with an interaction matrix to jointly capture local patterns and global interactions. The 1D convolution simulates neighborhood dependencies by aggregating local features through a sliding window, while the interaction matrix models the global relationships between drug and protein features via learned weights. This hybrid architecture preserves the spatial structure of pooled features and simultaneously enhances global semantic alignment across modalities. It can be formulated as:
\begin{equation}
\begin{aligned}
 & {{f}_{d}}=\text{Linear}(d_{graph}^{'}) \\
 & {{f}_{t}}=\text{Linear}(t_{graph}^{'}) \\
 & f_{d}^{'}=\text{GNN}({{f}_{d}})+\text{Softmax}(d_{graph}^{'}\cdot {{(t_{graph}^{'})}^{T}}){{f}_{d}}+{{f}_{d}} \\
 & f_{t}^{'}=\text{GNN}({{f}_{t}})+\text{Softmax}(t_{graph}^{'}\cdot {{(d_{graph}^{'})}}){{f}_{t}}+{{f}_{t}} \\
 & {{f}_{graph}}=\text{Linear}(f_{d}^{'}\|{f}_{t}^{'}) \\
\end{aligned}
\end{equation}
Where $\text{Linear}(\cdot)$ denotes the $\text{Linear}$ function, $\text{GNN}(\cdot)$ denotes the one-dimensional convolution function, $\text{Softmax}(\cdot)$ denotes the $\text{Softmax}$ function, and $\|$ denotes the join operation.

However, as the dimensionality of features increases and redundant information accumulates after intermodal interactions, traditional linear fusion becomes inadequate for compactly compressing the representation space and may introduce distracting noise. To overcome this limitation, we introduce a Deep Orthogonal Fusion (DOF) module \cite{su2024mski} at the late fusion stage (As shown in Fig.~\ref{fig2}(c)), aiming to refine the integration of cross-modal representations. The DOF module takes the representations from the three modalities as input and constructs diversity-enhanced features through linear transformation and orthogonal projection, thereby emphasizing modal discrepancies. A masking mechanism is then applied to filter out redundant components and reconstruct the informative ones, ultimately enhancing the discriminability and compactness of the fused representation. The formulation is as follows:
\begin{equation}
\begin{aligned}
& f_{t,g} = f_{txt} \| f_{graph}  \\
& f_{t,f} = f_{txt} \| f_{fun} \\
& f_{g,f} = f_{graph} \| f_{fun}  \\
& x_{12} = \left( f_{t,g} \| \left( f_{t,g} - \frac{f_{t,g} \cdot f_{t,f}}{\|f_{t,g}\|^2} f_{t,g} \right) \right) \\
& x_{13} = \left( f_{t,g} \| \left( f_{t,g} - \frac{f_{t,g} \cdot f_{g,f}}{\|f_{t,g}\|^2} f_{t,g} \right) \right) \\
& x_{23} = \left( f_{t,f} \| \left( f_{t,f} - \frac{f_{t,f} \cdot f_{g,f}}{\|f_{t,f}\|^2} f_{t,f} \right) \right)  \\
& x_{\text{Red}} = \text{Linear}(x_{12} \| x_{13} \| x_{23}) \\
& \text{mask}_m = \mathbb{I}(\text{sim} > \text{threshold})
\end{aligned}
\end{equation}
Where ${{f}_{later}}\in {{\mathbb{R}}^{Seq\times Dim}}$ is the drug-target interaction feature representation after multi-modal late fusion. ${{x}_{12}}, {{x}_{13}}, {{x}_{23}}, {{x}_{\text{Red}}}$ is the redundancy of the orthogonal calculation after said, $\mathbb{I}(\cdot)$ is the indicator function, $\text{threshold}$ is a threshold, It is used to filter out the most similar targets, avoid redundancy and retain specific information. $\text{mask}_m$ is to calculate the similarity between the redundant representation and the drug-target interaction feature representation of different modalities,
and generate a binary mask based on the similarity score. Like attention mechanism, we use key/value pair ($ {x}_{\text{Red}}, {{f}_{m}}$) and query vector ${{f}_{m}}$ to calculate the cosine similarity:
\begin{equation}
\begin{aligned}
 & \text{sim}_{m} = \text{Cos}(tK({{x}_{\text{Red}}}),tQ({{f}_{m}}))\  \\
 & {{z}_{m}}=\text{mask}_{m}\times tV({{f}_{m}})\ ,\ m=\text{txt}/\text{graph}/\text{fun} \\
 & {{f}_{later}}=\text{Linear}({{z}_{txt}}\|{{z}_{graph}}\|{{z}_{fun}}\|{{x}_{\text{Red}}}) \\
\end{aligned}
\end{equation}
Where $tQ(\cdot),tK(\cdot),tV(\cdot)$ denote the linear transformation. Guided by the estimated similarity $\text{sim}$, we filter out features that are relatively similar (high similarity scores) and only select features of a particular modality (low similarity scores) for the final fusion.

\subsubsection{Interaction Prediction}
The proposed $M^3$ST-DTI method is a multi-task learning framework. We integrate early fusion features and late fusion features through linear connections:
\begin{align}
 & f_{\text{output}} = \text{Linear}(f_{\text{early}}\|f_{\text{later}})
\end{align}
We use different loss functions to train 6 branching tasks including: ${{\text{CE}}_{\text{txt}}}$, ${{\text{CE}}_{\text{graph}}}$, ${{\text{CE}}_{\text{fun}}}$, ${{\text{CE}}_{\text{early}}}, {{\text{CE}}_{\text{later}}}, {{\text{CE}}_{\text{output}}}$, so, our total classification loss:
\begin{equation}
\begin{aligned}
& L_c = \frac{L_{\text{modal}} + L_{\text{stage}} + \text{CE}_{\text{output}}}{3}  \\
& L_{\text{modal}} = \frac{\text{CE}_{\text{txt}} + \text{CE}_{\text{graph}} + \text{CE}_{\text{fun}}}{3}  \\
& L_{\text{stage}} = \frac{\text{CE}_{\text{early}} + \text{CE}_{\text{later}}}{2}
\end{aligned}
\end{equation}
Our alignment loss Gram Loss includes drug feature alignment loss $\text{Gram}_{d}$ and target feature alignment loss $\text{Gram}_{t}$. The total alignment loss is as follows:
\begin{align}
 & {{L}_{g}}=(\text{Gram}_{d}+\text{Gram}_{t})/2
\end{align}
In the end, we obtain the final loss function as follows:
\begin{align}
 & L = ({{L}_{c}} + \lambda {{L}_{g}})/(1 + \lambda)
\end{align}
Where $\lambda$ is the loss weight.

\section{Results}
\subsection{Experimental Settings}
The initialization parameters of $M^3$ST-DTI are configured as follows: the training batch size is set to 32; the Adam optimizer \cite{kingma2014adam} is used with a learning rate of 1e-4 and a weight decay coefficient of 1e-5. The learning rate decays every 10 epochs. The maximum number of training epochs is 50. The hidden layer dimension is set to 512. The number of layers in the graph self-attention module is 3, and the number of layers in the multi-head self-attention module is 1. The loss weight is set to 1.0, and the temperature parameter is 0.1.

\subsection{Ablation Experiment}
To comprehensively assess the contribution of each module in the proposed $M^3$ST-DTI framework, we performed ablation experiments on the BindingDB dataset. Specifically, we examined the roles of four core components: multi-modal feature extraction, early fusion, late fusion, and the interaction-aware prediction head. Ablation settings include three single-modality variants (text-only, graph-only, function-only), two fusion strategies (early-only, late-only), and the full $M^3$ST-DTI model. The results of the ablation study are presented in Table~\ref{tab2}.
\begin{table}[htbp]
\caption{Performance on BindingDB Dataset (\% Metrics)}
\begin{center}
\begin{tabular}{|l|c|c|c|c|}
\hline
\textbf{Model} & \textbf{Accuracy} & \textbf{F1-Score} & \textbf{AUROC} & \textbf{AUPRC} \\
\hline
Only txt & 81.92 & 56.56 & 88.46 & 56.24 \\
Only graph & 80.00 & 55.37 & 89.59 & 55.27 \\
Only fun & 81.45 & 57.10 & 87.14 & 55.46 \\
Only early & 82.42 & 58.69 & 90.25 & 56.44 \\
Only later & 83.05 & 59.26 & 90.83 & 59.09 \\
Ours & 83.36 & 60.38 & 91.04 & 60.47 \\
\hline
\end{tabular}
\end{center}
\label{tab2}
\end{table}

Multimodal fusion provides clear benefits for DTI prediction. On BindingDB, text modality (Only txt) performs the best among single modalities, but there is still a large room for improvement, confirming the necessity of cross-modal integration. The fusion strategy analysis shows that the only early variant improves over all unimodal baselines. However, its performance is lower than that of the full model. In contrast, the only later variant, with BCA and DOF modules, achieved a higher f1 value, confirming the benefit of deep interactive fusion. However, without early alignment, the overall performance is still suboptimal. The full $M^3$ST-DTI framework, which integrates multimodal feature extraction, semantic alignment, front-back fusion, and interactive perceptual prediction, achieves the best results on all metrics. These results validate the effectiveness of the two-stage fusion design in capturing coarse-to-fine semantic relationships, highlighting the robustness of the multimodal, multi-stage approach.
\subsection{Comparative Experiment}
\begin{figure}[htbp]
\centerline{\includegraphics[width=\columnwidth]{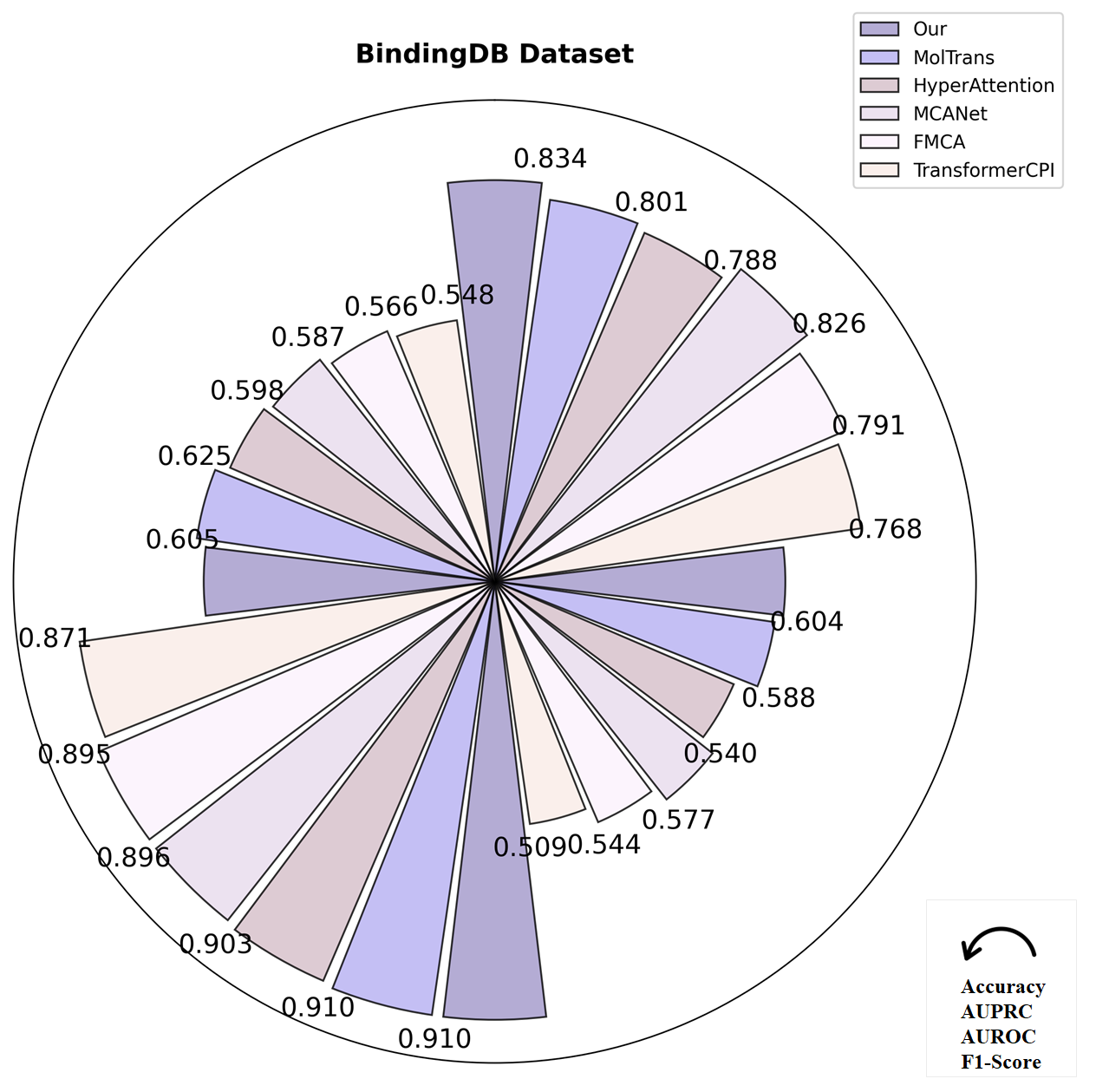}}
\caption{Radial Plot of the Comparison Experiment Results}
\label{fig4}
\end{figure}

To comprehensively assess the effectiveness of $M^3$ST-DTI, we conducted comparative experiments on the BindingDB dataset against several leading baseline models. As shown in Fig.~\ref{fig4}, our model consistently achieved superior performance across multiple evaluation metrics, including Accuracy, F1-score, AUROC, and AUPRC, demonstrating its strength in both representation learning and multi-modal feature alignment.

On the BindingDB dataset, $M^3$ST-DTI outperformed strong baselines such as MolTrans, MCANet, HyperAttention, and FMCA, achieving an Accuracy of 83.36\%, an F1-score of 60.47\%, and an AUROC of 91.04\%. Notably, the model exhibited greater robustness in terms of F1-score, indicating enhanced generalization under imbalanced sample distributions. While MolTrans achieved a slightly higher AUPRC, our model maintained the most balanced overall performance, effectively identifying positive interactions while preserving a high recall rate.

\subsection{Visual Analysis}
\begin{figure}[htbp]
\centerline{\includegraphics[width=0.5\textwidth]{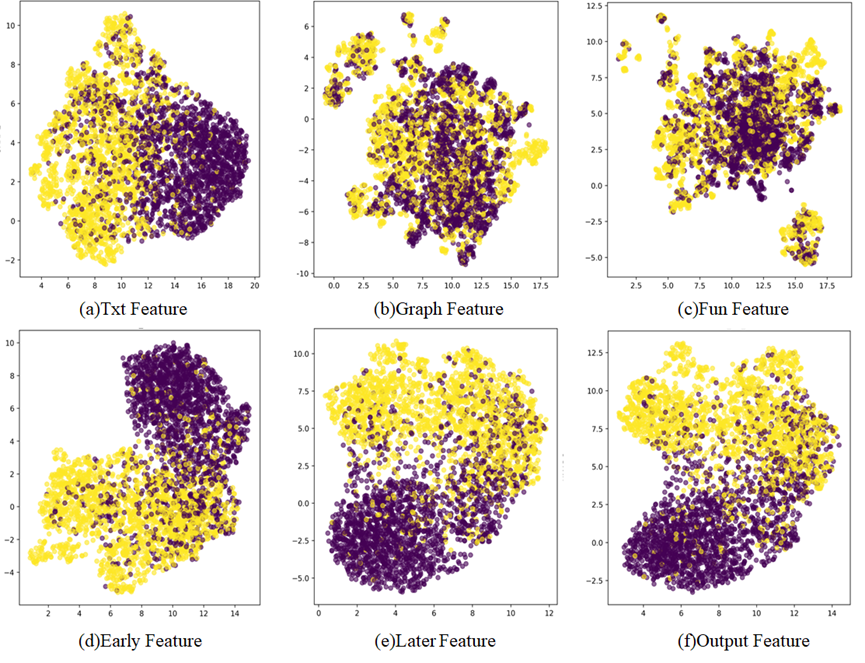}}
\caption{UMAP Visualization of feature distributions}
\label{fig5}
\end{figure}
To gain a deeper understanding of the model performance in multi-modal feature extraction and fusion, the dimensionality reduction visualization of features from different modalities and fusion stages is performed based on umap. As shown in Fig.~\ref{fig5}, in the early fusion stage (d), under the joint influence of MCA and Gram alignment loss, the embedding space becomes more structured with obvious intra-class compactness and inter-class discreteness. In the late fusion (e), the addition of BCA and DOF modules further sharps the class boundaries, and the concentration between negative samples increases. The final output feature (f) exhibits an optimal clustering structure where the boundary between positive and negative classes becomes clear and class overlap is greatly reduced. These results show that the model progressively optimizes the feature representation throughout the fusion stage. Our multi-stage fusion strategy effectively captures complex molecular-protein interaction patterns, continuously enhancing the discriminability from raw single-modal input to the final joint representation.
\section{Conclusion}
This study introduces $M^3$ST-DTI, a novel framework that integrates multi-modal feature representation with multistage alignment strategies, achieving marked advances in drug-target interaction (DTI) prediction.    Experimental results on the BindingDB dataset strongly validate the superior performance of $M^3$ST-DTI.    Moreover, UMAP-based visualizations reveal a layer-wise progression in feature optimization from unimodal inputs to the final representations, showing increasingly distinct boundaries between positive and negative samples in the embedding space and significantly enhanced class separability. Future work will extend $M^3$ST-DTI's functionality and explore its broader biomedical applications to enhance drug development and therapeutic research.

\vspace{12pt}
\bibliographystyle{IEEEtran}
\bibliography{myref}
\end{document}